\documentclass[aps,prb,twocolumn,showpacs,superscriptaddress]{revtex4-1}
\usepackage{graphicx}
\usepackage{dcolumn}
\usepackage{bm}
\usepackage{amssymb}
\usepackage{amsmath}
\usepackage{subfigure}
\usepackage{float}
\begin{document}
\title{Momentum-Resolved Electronic Relaxation Dynamics in D-wave Superconductors}
\author{Yong Wang}
\affiliation{Department of Physics, The University of Hong Kong, Hong Kong SAR, China}
\author{Fu-Chun Zhang}
\affiliation{Department of Physics, The University of Hong Kong, Hong Kong SAR, China}
\affiliation{Department of Physics, Zhejiang University, China}
\begin{abstract}
Motivated by recent development in time-resolved angle-resolved photoemission spectroscopy (trARPES) for d-wave superconductors, we analyze the non-equilibrium relaxation dynamics of the laser pulse excited sample within the scenario of two-temperature model. It is shown that the main features reported in the trARPES technique may be understood within this phenomenological picture. The momentum dependence of the excited quasiparticle density and the relaxation rate is associated with the dynamics of the nodal d-wave superconducting gap, and the fluence dependence of the relaxation rate is related to the recombination process of quasiparticles into Cooper pairs.
\end{abstract}
\maketitle
\section{Introduction} Study of temporal dynamics has greatly advanced our understanding of correlated electron systems for such dynamics may provide information about essential energy scales associated with the elementary excitations and collective modes in these systems.\cite{rev} The newly developed time-resolved angle-resolved photo-emission spectroscopy (trARPES) is a powerful technique to directly measure momentum resolved electronic dynamics,\cite{exp0} and has been applied to study ultrafast dynamics in charge density wave materials,\cite{cdw1,cdw2} high-temperature superconductors(HTSCs),\cite{sc1,sc2,sc3} and topological insulators.\cite{ti} Smallwood \emph{et al.} ~\cite{sc3} have used the technique to study gap and quasiparticle (QP) population dynamics in optimally doped d-wave HTSC Bi$_2$Sr$_2$CaCu$_2$O$_{8+\delta}$. Their results clearly show that QP's relaxation rate is dependent on the momentum and fluence.  These intriguing observations have challenged the understanding of non-equilibrium dynamics in the superconducting (SC) state, and called for theoretical explanations for further explorations on HTSCs with trARPES technique. In this paper, we apply a two-temperature model~\cite{exp0} to theoretically study the transient energy distribution and QP relaxation dynamics of d-wave superconductor measured in trARPES. Our theory explains the momentum-resolved dynamics of the
photoexcited QPs, and is in good agreement with experiments.

\section{Theoretical Model}
\subsection{Two-Temperature Scenario}
In a typical trARPES experiment, a pump laser pulse is first applied to excite the investigated sample into a non-equilibrium state, and ARPES technique is then used to measure the temporal electronic dynamics of the sample. Presently, trARPES technique has a time-resolution $\tau_{ARPES} \approx 100\sim 300$ fs.\cite{sc1,sc2,sc3} In the relaxation process of the laser-excited d-wave superconductors, there are two essential time scales determined by the intrinsic interactions, namely, $\tau_{ee}\sim\mathcal{O}(\text{fs})$ due to the electron-electron (e-e) scattering and $\tau_{ep}\sim\mathcal{O}(\text{ps})$ due to the electron-phonon (e-p) scattering. We consider a typical case where $\tau_{ee}\ll \tau_{ARPES} \ll \tau_{ep}$, which is suitable for the present experiment resolution of $\tau_{ARPES}$. In this case, the e-e scattering thermalizes the excited electronic subsystem into a quasi-equilibrium state in the time scale of $\tau_{ee}$. Then the detailed non-equilibrium processes of the electronic subsystem will be washed out during the time $\tau_{ARPES}$, and the quasi-equilibrium state can be characterized by the effective temperature $T_{e}$. In the longer time scale $\tau_{ep}$, the extra energy in the electronic subsystem will be dissipated to the lattice subsystem through electron-phonon interaction, and $T_{e}$ will decrease to the lattice temperature $T_{l}$. Thus the observed gap and QP dynamics of the superconductors will be associated with the time-dependent electronic temperature $T_{e}$. This simple two-temperature scenario has been applied to describe the relaxation dynamics in HTSC, \cite{exp0} and will be applied to study the trARPES here.

The dependence of $T_{e}$ on the time $t$ in the two-temperature model is not known \emph{a priori}, but may be determined from the experimental measurements or derived from the microscopic theory. For simplicity, we assume that the dominant process for the electronic system to dissipate energy is the pairwise recombination of QPs into Cooper pairs, and the Rothwarf-Taylor equation\cite{RT} for the total QP number still holds, which is supported by the previous THz conductivity measurement.\cite{exp1} Further exploiting the fact that the total QP density is approximately proportional to the square of electronic temperature $T_{e}$, then the equation of the electronic temperature $T_{e}$ is estimated as\cite{TeEq}
\begin{eqnarray}
\frac{dT_{e}^{2}}{dt}=-r(T_{e}^{4}-T_{l}^{4}).\label{decay}
\end{eqnarray}
Here, the pump laser is assumed to be applied at time $t=0$, which promotes the electronic temperature from $T_{l}$ to $T_{e}(0)$. For given $T_{l}$, higher laser fluence will drive the sample into higher excited state and lead to larger $T_{e}(0)$.  Then Eq.~(\ref{decay}) describes the decay of the electronic temperature from $T_{e}(0)$ to $T_{l}$, and the parameter $r$ is proportional to the recombination rate of QPs\cite{TeEq} and will be determined by fitting to the experimental results later on.

\begin{figure}[htbp]
\includegraphics[scale=0.16,clip]{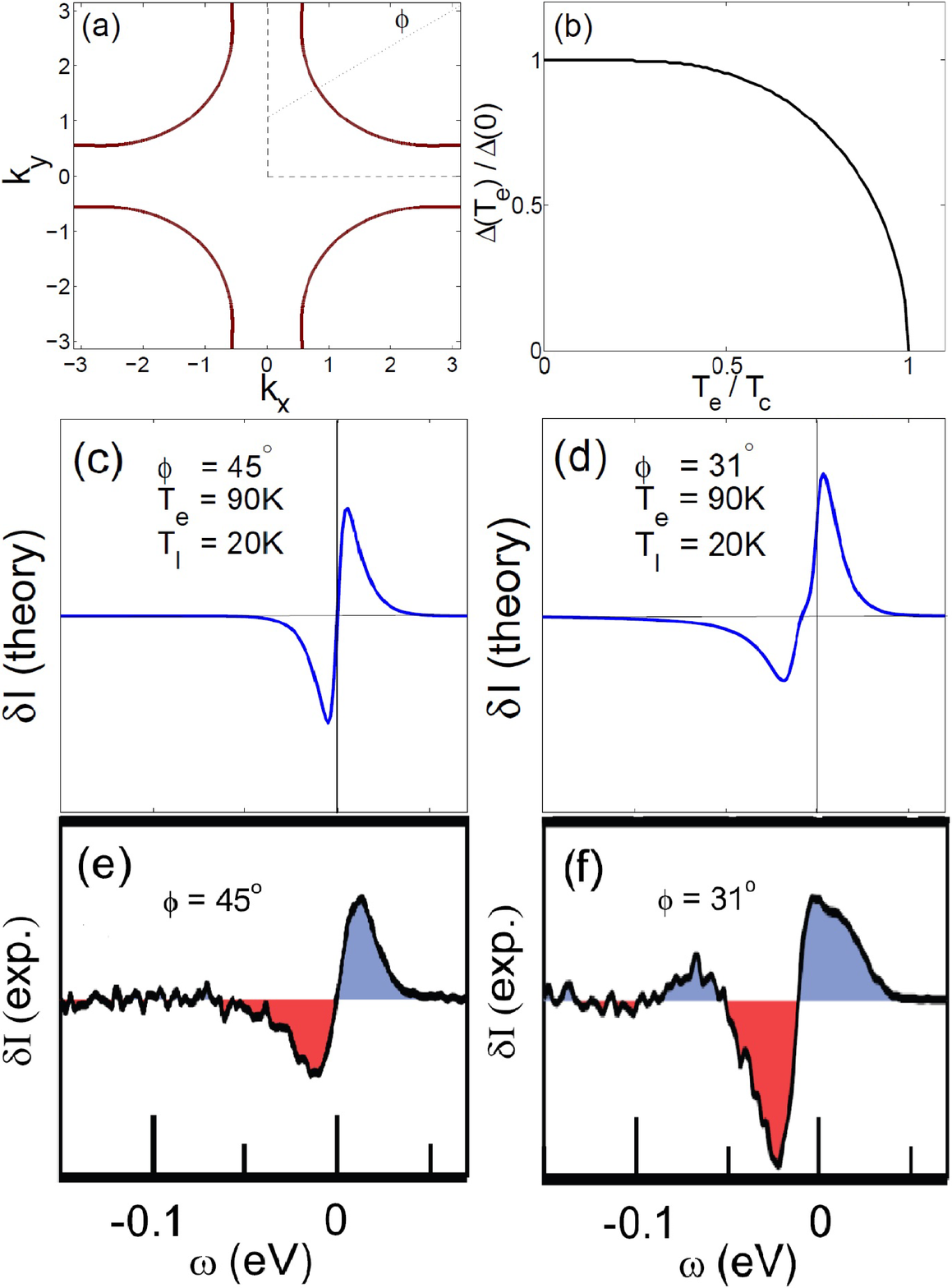}
\caption{(Color online) (a) Fermi surface in a tight-binding model\cite{tb} studied in the present paper for optimally-doped Bi-2212. $\phi$ denotes the angle to describe momentum cutline in ARPES measurements. (b) Temperature dependence of SC gap, determined from Eq.~(\ref{gap}). (c) and (d): Calculated change in the line-momentum-integrated ARPES intensity, $\delta I(\omega)$, defined in Eq.~(\ref{deltaI}), between $T_{e}$=90K and $T_{l}$=20K, along the momentum cutlines $\phi=45^{\circ}$ and $\phi=31^{\circ}$, respectively. (e) and (f): Experimental result of the intensity change $\delta I(\omega)$ along the momentum cutlines $\phi=45^{\circ}$ and $\phi=31^{\circ}$, respectively,  taken from Ref.~\onlinecite{sc3} for pump fluence 5~$\mu$J/cm$^{2}$. Blue color represents intensity gain and red color intensity loss. }\label{bandgap}
\end{figure}

\subsection{Band Structure, Gap Function, Spectral Function, ARPES Intensity}
Now we consider the quasi-equilibrium electronic state characterized by temperature $T_{e}$ and the resulting consequences in trARPES measurements. Since the trARPES has been reported on the optimally-doped Bi$_{2}$Sr$_{2}$CaCu$_{2}$O$_{8+\delta}$(Bi-2212) samples,\cite{sc1,sc2,sc3} our calculations below will be specified to this compound although the main conclusions are expected to apply to general d-wave superconductors. We use a simple tight-binding model suggested in Ref.~\onlinecite{tb} to describe the normal state energy dispersion, with
\begin{eqnarray}
\epsilon_{\mathbf{k}}&=&c_{0}+\frac{c_{1}}{2}(\cos k_{x}+\cos k_{y})+c_{2}\cos k_{x}\cos k_{y}\nonumber\\
&-&\frac{c_{3}}{2}(\cos 2k_{x}+\cos 2k_{y})+c_{5}\cos 2k_{x}\cos 2k_{y}\nonumber\\
&+&\frac{c_{4}}{2}(\cos 2k_{x}\cos 2k_{y}+\cos k_{x}\cos k_{y}),\label{band}
\end{eqnarray}
where the coefficients are $c_{0}=0.1305,c_{1}=-0.5951,c_{2}=0.1636,c_{3}=-0.0519,c_{4}=-0.1117,c_{5}=0.0510$.
The Fermi surface (FS) determined by $\epsilon_{\mathbf{k}}=0$ is shown in Fig.~\ref{bandgap}(a). A momentum cutline denoted by an angle $\phi$ for ARPES measurements is also shown.

We consider a simple d-wave SC gap function for Bi-2212, $\Delta_{\mathbf{k}}=\frac{\Delta(T_{e})}{2}(\cos k_{x}-\cos k_{y})$, and further assume a simple BCS form for the temperature dependence of $\Delta(T_{e})$, which is determined by the self-consistent equation\cite{Tinkham}
\begin{eqnarray}
\frac{1}{N(0)V}=\int_{0}^{\hbar\omega_{c}}\frac{\text{tanh}[\frac{1}{2k_{b}T_{e}}(\xi^{2}+\Delta^{2})^{1/2}]}{(\xi^{2}+\Delta^{2})^{1/2}}d\xi.\label{gap}
\end{eqnarray}
Here, $k_{b}$ is the Boltzmann constant, $N(0)$ denotes the density of states at the Fermi level of one spin orientation, $V$ characterizes the strength of pair potential for the electrons within the cutoff energy $\hbar\omega_{c}$.\cite{Tinkham} With the weak-coupling condition $\hbar\omega_{c}\gg k_{b}T_{c}$ and the relation $\frac{\Delta(0)}{k_{b}T_{c}}=1.764$, where $T_{c}$ is the SC critical temperature, Eq.~(\ref{gap}) gives the universal function of $\Delta(T_{e})/\Delta(0)$ on $T_{e}/T_{c}$,\cite{Tinkham} as shown in Fig.~\ref{bandgap}(b). While this form is for conventional s-wave superconductor, the deviation from d-wave superconductors is not significant to change the qualitative results below.

With the bare band dispersion relation $\epsilon_{\mathbf{k}}$ and the gap function $\Delta_{\mathbf{k}}$, the QP excitation energy $E_{\mathbf{k}}$ is determined as $E_{\mathbf{k}}=\sqrt{\epsilon_{\mathbf{k}}^{2}+\Delta_{\mathbf{k}}^{2}}$. The key quality measured in the trARPES is the spectral function $A(\omega,\mathbf{k})$ in terms of the energy distribution curves, which takes the form below with a Lorentzian lineshape\cite{spec}
\begin{eqnarray}
A(\omega,\mathbf{k})=\frac{1}{\pi}[\frac{\mu_{\mathbf{k}}^{2}\Gamma}{(\omega-E_{\mathbf{k}})^{2}+\Gamma^{2}}+\frac{\nu_{\mathbf{k}}^{2}\Gamma}{(\omega+E_{\mathbf{k}})^2+\Gamma^{2}}].\label{specfuc}
\end{eqnarray}
Here, $\Gamma$ characterizes the broadening of the spectral line and is assumed to be a constant. $\nu_{\mathbf{k}}^{2}=1-\mu_{\mathbf{k}}^{2}=\frac{1}{2}(1-\epsilon_{\mathbf{k}}/E_{\mathbf{k}})$ are the coherence factors. In our calculations, we set $T_{c}=90$~K, $\Delta(0)=0.03$~eV, and $\Gamma=0.01$~eV. When the electronic temperature $T_{e}$ changes, the spectrum function $A(\omega,\mathbf{k})$ also changes due to the temperature-dependent SC gap.

Several physical qualities based on the spectral functions are usually given in the trARPES experiments, and will be calculated below in order to compare with the experimental results. One is the line-momentum-integrated ARPES intensity $I(\omega)$ (ARPES intensity hereafter)  along a momentum cutline $L$ (the choice in our calculations is shown in Fig.~\ref{bandgap}(a)),
\begin{eqnarray}
I(\omega) =f(\omega) \int_{L}d\mathbf{k}A(\omega,\mathbf{k}),\label{inten}
\end{eqnarray}
where $f(\omega)$ is the Fermi-Dirac (FD) distribution function, which depends on the electronic temperature $T_{e}$. Note that $A(\omega,\mathbf{k})$ is also a function of $T_{e}$ via the SC gap function. In the two-temperature scenario, as $T_{e}$ decreases from $T_{e}(0)$ to the equilibrium temperature $T_{l}$, the SC gap increases and the thermal distribution of QPs is suppressed, then the number of excited QPs is reduced. This process is reflected in the time-evolution of the  ARPES intensity $I(\omega)$.\cite{sc1,sc2,sc3}

A more relevant quantity is the change of the ARPES intensity between the two temperatures $T_{e}$ and $T_{l}$,
\begin{eqnarray}
\delta I(\omega)&\equiv&I(\omega;T_{e})-I(\omega;T_{l}),\label{deltaI}
\end{eqnarray}
It is convenient to introduce the line-momentum-integrated density of states(DOS hereafter) along the momentum cutline $L$, $D_{L}(\omega)=\int_{L}d\mathbf{k}A(\omega,\mathbf{k})$, then  $\delta I(\omega)$ is decomposed into two terms,
\begin{eqnarray}
\delta I(\omega)=D_{L}(\omega;T_{l})\delta f(\omega)+\delta D_{L}(\omega)f(\omega;T_{e}),\label{diffint}
\end{eqnarray}
where $\delta f(\omega)\equiv f(\omega;T_{e})-f(\omega;T_{l})$ is the change of the distribution function, and $\delta D_{L}(\omega)\equiv D_{L}(\omega;T_{e})-D_{L}(\omega;T_{l})$ is the change of the DOS along $L$. This decomposition is vital for us to understand the results below. In Fig.~\ref{bandgap}(c) and (d), we show the results of $\delta I(\omega)$ for the system for two sets of FS angles, i.e. a diagonal cutline $\phi=45^{\circ}$ (nodal cutline) and an off-nodal cutline $\phi=31^{\circ}$, respectively. The two temperatures are $T_{e}=90$K, and $T_{l}=20$K. $\delta I$ is strongly dependent on the cutline.  For the nodal cutline, $\delta I$ is approximately antisymmetric with respect to $\omega$, similar to the change of the Fermi distribution function $\delta f(\omega)$. In particular, $\delta I =0$ at $\omega=0$.  For the non-nodal cutline, the shape is far from the antisymmetric, and $\delta I(\omega=0)$ is finite and positive, and $\delta I=0$ occurs at $\omega < 0$.  Our calculations are in good agreement with the experimental data,\cite{sc3} which are reproduced in Fig.~\ref{bandgap}(e) and (f) for comparison. The strong FS angle dependence of the ARPES intensity is associated with the symmetry of the SC gap function. For the nodal cutline of ($\phi=45^{\circ}$), the second term in Eq.~(\ref{diffint}) vanishes. Furthermore, $\delta f(-\omega)=-\delta f(\omega)$ and $\delta f(\omega)$ is of substantial value only within a window of $|\omega| < 3k_{b}T_{e}$. Since $D_{L}(\omega;T_{l})$ doesn't vary sharply near $\omega=0$, we have $\delta I(\omega)$ approximately antisymmetric as numerically shown in Fig.~\ref{bandgap}(c). For the off-nodal case, the second term in Eq.~(\ref{diffint}) plays an important role, and $\delta I(\omega)$ is no longer antisymmetric.  At $\omega=0$, we have  $\delta D_{L}(0)>0$ according to Eq.~(\ref{specfuc}) and $f(0)=\frac{1}{2}$, thus $\delta I(0)=\frac{1}{2}\delta D_{L}(0)>0$.

Another important quality given in trARPES experiments is the angle-resolved QP density. Following the experiments,\cite{sc1,sc3} we define $I$ as the integral of the ARPES intensity $I(\omega)$ over $\omega$ above the Fermi energy along the momentum cutline $L$ at temperature $T_{l}$, and $\Delta I$ is the change of $I$ when the electronic temperature increased from $T_{l}$ to $T_{e}$, i.e.
\begin{eqnarray}
I=\int_{0}^{\infty}d\omega I(\omega),\quad\Delta I=\int_{0}^{\infty}d\omega \delta I(\omega).\label{IdI}
\end{eqnarray}
Below we analysis the dynamics of $\Delta I$ to further explain the trARPES experiments.
\section{Calculation Results}
\subsection{Angle Dependence of Photoexcited Quasiparticle Density} 
\begin{figure}[htbp]
\includegraphics[scale=0.16,clip]{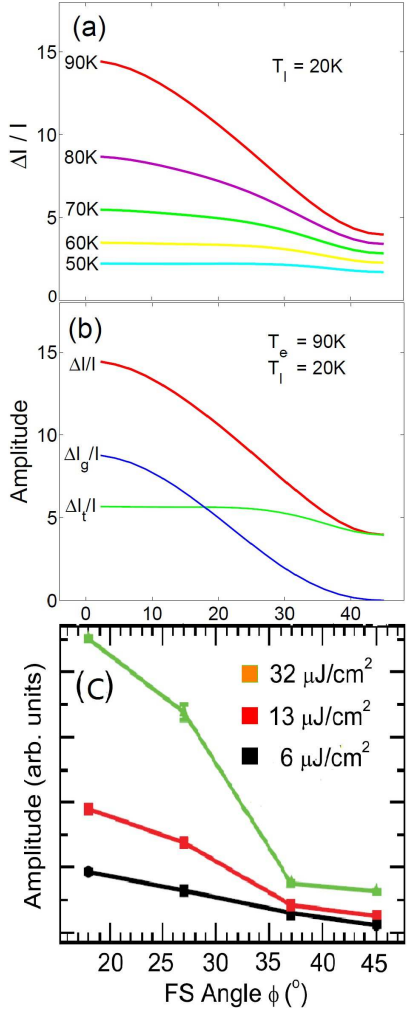}
\caption{(Color online)(a) The dependence of relative ARPES intensity change $\Delta I/I$ on the Fermi surface angle $\phi$ for several temperatures $T_{e}$ with $T_{l}=20$K. (b) Thermal broadening  contribution $\Delta I_{t}/I$ and gap reduction contribution $\Delta I_{g}/I$ to the relative ARPES intensity change $\Delta I/I$ for temperature $T_{e}=90$K and $T_{l}=20$K. (c) Experimental result for the photoexcited QP density, taken from Ref.~\onlinecite{sc1}. }.\label{IniInt}
\end{figure}

It has been observed that the photoexcited QP density is strongly dependent on the FS angle $\phi$ similar to the d-wave SC gap function.\cite{sc1} According to our discussions above, the electronic temperature has been increased from $T_{l}$ to $T_{e}(0)$ after the laser pumping, because the detailed electronic dynamics to reach such a quasi-equilibrium state has been washed out due to the large time resolution of the ARPES measurements. Then the SC gap is reduced and even closed if $T_{e}(0)\geq T_{c}$ and more QPs are also excited. Thus we can calculate the photoexcited QP density according to Eq.~(\ref{IdI}), and the results of $\Delta I/I$ for the temperatures $T_{e}(0)$=90K, 80K, 70K, 60K, 50K and $T_{l}$=20K are given in Fig.~\ref{IniInt}(a). The calculation results reproduce the angle-dependence relation observed in experiments,\cite{sc1} as shown in Fig.~\ref{IniInt}(c). Notice that higher pump fluence corresponds to higher electronic temperature $T_{e}(0)$.

To understand the results, we recall that $\Delta I$ has both the contributions from thermal broadening and gap reduction according to Eq.~(\ref{diffint}). We denote $\Delta I\equiv\Delta I_{t}+\Delta I_{g}$, where
$\Delta I_{t}=\int_{0}^{\infty}d\omega D_{L}(\omega;T_{l})\delta f(\omega),\quad
\Delta I_{g}=\int_{0}^{\infty}d\omega \delta D_{L}(\omega)f(\omega;T_{e}).$
The angle dependence of $\Delta I_{t}/I$ is expected to be weak as long as $D_{L}(\omega,T_{l})$ doesn't vary sharply near the Fermi energy. Specially, if $D_{L}(\omega;T_{l})$ is approximated by its value $D_{L}(0;T_{l})$ at Fermi energy, $\Delta I_{t}/I$ will be independent on the FS angle $\phi$. While for $\Delta I_{g}/I$, its angle dependence comes from the change of DOS $\delta D_{L}(\omega)$ which is strongly dependent on the d-wave gap function. For the nodal case($\phi=45^{\circ}$), $\delta D_{L}(\omega)=0$ gives $\Delta I_{g}/I=0$; while in the off-nodal region, the gap reduction will increase the DOS near the Fermi energy and thus $\Delta I_{g}/I>0$. As an example, Fig.~\ref{IniInt}(b) shows the two contributions $\Delta I_{t}/I$ and $\Delta I_{g}/I$ to $\Delta I/I$ for $T_{e}(0)=90$K and $T_{l}=20$K. It is found that $\Delta I_{t}/I$ is only weakly dependent on the FS angle $\phi$, while $\Delta I_{g}/I$ strongly depends on $\phi$ in the similar way as the d-wave gap function, which supports our analysis. When the pump fluence is lower, the temperature $T_{e}(0)$ is closer to $T_{l}$, then the gap reduction becomes smaller, and the angle dependence of $\Delta I/I$ becomes weak. This tendency is shown in Fig.~\ref{IniInt}(a), and is also found in the trARPES experiments in Fig.~\ref{IniInt}(c).

\subsection{Angle Dependence of Quasiparticle Decay Rate}
\begin{figure}[htbp]
\includegraphics[scale=0.16,clip]{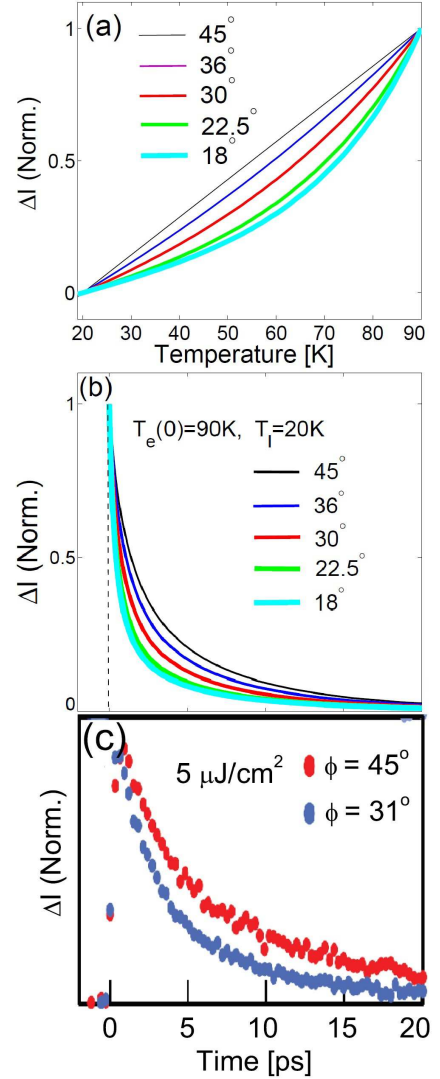}
\caption{(Color online)(a) The temperature dependence of normalized $\Delta I$ for several FS angles with initial electronic temperature $T_{e}(0)$=90K and lattice temperature $T_{l}$=20K. (b) Decay curves of normalized $\Delta I$ for several FS angles by mapping the curves in (a) to the time scale in the help of Eq.~(\ref{decay}). The fitting parameter $r=1.5\times10^{-4}\text{K}^{-2}\text{ps}^{-1}$. (c) Experimental decay curves of normalized $\Delta I$ at $\phi=45^{\circ}$ and $\phi=31^{\circ}$ respectively, taken from Ref.~\onlinecite{sc3}. }\label{TempScale}
\end{figure}

We further analyze the angle-dependent decay rate of QPs. It has been observed that the QPs in the off-nodal region decay faster than the ones in the nodal region.\cite{sc3} In the two-temperature scenario, the ARPES measures the quasi-equilibrium electronic state characterized by the temperature $T_{e}(t)$. We first calculate the normalized $\Delta I$ for different temperatures at given FS angle $\phi$, and the results are shown in Fig.~\ref{TempScale}(a). Here, we have fixed $T_{l}=20$K, and varied the temperature $T_{e}$ from $90$K to $20$K, and then normalize $\Delta I$ by its value at $90$K. It is found that for given temperature $T_{e}$, the normalized $\Delta I$ decreases when the angle $\phi$ moves from the nodal region to the off-nodal region. Then with the help of Eq.~(\ref{decay}), the temperature dependence of the normalized $\Delta I$ in Fig.~\ref{TempScale}(a) is mapped to the corresponding time-dependence relation, as shown in Fig.~\ref{TempScale}(b). Here we have set $r=1.5\times10^{-4}~\text{K}^{-2}\text{ps}^{-1}$ to fit the experimental results. Our results here reproduce the observations that the decay of normalized $\Delta I/I$ depends on the FS angle\cite{sc3} shown in Fig.~\ref{TempScale}(c).

The results in Fig.~\ref{TempScale} can also be understood from the two contributions $\Delta I_{t}$ and $\Delta I_{g}$ in $\Delta I$. For the nodal case ($\phi=45^{\circ}$), we have $\Delta I_{g}=0$, while $\Delta I_{t}\sim D_{L}(0;T_{l})k_{b}(T_{e}-T_{l})$ if the DOS $D_{L}(\omega;T_{l})$ is approximated by its value at Fermi energy. This explains the nearly linear dependence of the normalized $\Delta I$ on the temperature for $\phi=45^{\circ}$ in Fig.~\ref{TempScale}(a). In the off-nodal case, $\Delta I_{g}$ due to the gap variation contributes to $\Delta I$ in addition to $\Delta I_{t}$. When the temperature $T_{e}$ decreases, the gap becomes larger and $\delta D_{L}(\omega)$ decreases. Thus in the off-nodal region the normalized $\Delta I$ deviates from the linear temperature-dependence relation and decays faster, as shown in Fig.~\ref{TempScale}(a). After mapping to the time scale, the normalized $\Delta I$ then shows the angle-dependent decay rate of QPs.
\begin{figure}[htbp]
\includegraphics[scale=0.16,clip]{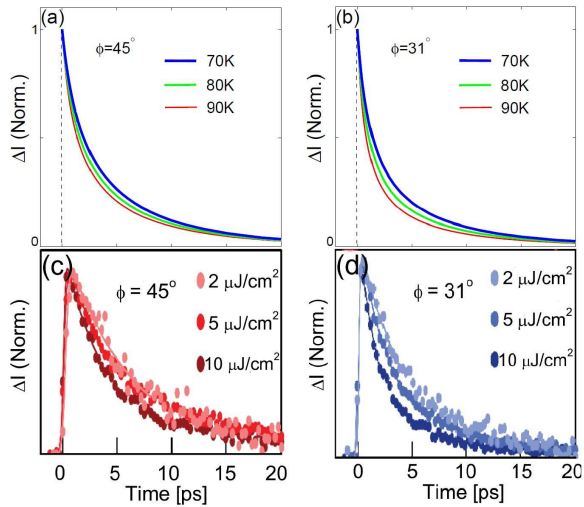}
\caption{(Color online)(a) and (b): Decay curves of normalized $\Delta I$ with different initial electronic temperature $T_{e}(0)$ and lattice temperature $T_{l}=20$K for $\phi=45^{\circ}$ and $\phi=31^{\circ}$ respectively. The fitting parameter $r=1.5\times10^{-4}\text{K}^{-2}\text{ps}^{-1}$ in Eq.~(\ref{decay}). (c) and (d) : Experimental decay curves of normalized $\Delta I$ with different pumping fluences for $\phi=45^{\circ}$ and $\phi=31^{\circ}$ respectively, taken from Ref.~\onlinecite{sc3}. Higher pumping fluence corresponds to higher initial electronic temperature $T_{e}(0)$.}
\label{FluScale}
\end{figure}
\subsection{Fluence Dependence of Quasiparticle Decay Rate}
It is also observed that the QPs at fixed angle relax faster if the sample is pumped by higher fluence laser.\cite{sc3} The reason however is not the same as the angle-dependence case discussed above, considering that the superconducting gap is not involved in the nodal direction ($\phi=45^{\circ}$). The nearly linear dependence of $\Delta I$ on the temperature $T_{e}$ shown in Fig.~\ref{TempScale}(a) implies that the decay of the normalized $\Delta I$ along the cutline $\phi=45^{\circ}$ directly reflects the decay of $T_{e}$. This fluence-dependent decay behavior can be explained by the recombination process of QPs into Cooper pairs,\cite{exp1} and thus justifies the assumed decay equation (\ref{decay}) for $T_{e}$. With Eq.~(\ref{decay}) and the parameter $r=1.5\times10^{-4}~\text{K}^{-2}\text{ps}^{-1}$, we got the decay of the normalized $\Delta I$ with different initial temperatures $T_{e}(0)$ for $\phi=45^{\circ}$ and $\phi=31^{\circ}$, as shown in Fig.~\ref{FluScale}(a) and (b). The calculation results reproduces the experimental observations that higher fluence induces faster QP decay rate, as shown in Fig.~\ref{FluScale}(c) and (d). Thus, unlike the angle-dependent decay rate of QPs which is associated with the d-wave gap dynamics, the fluence-dependent decay rate of QPs is due to the pairwise recombination of QPs into Cooper pairs.

\section{Conclusion}
In conclusion, we show that the main features of the trARPES observations for d-wave superconductors so far can be explained within a simple two-temperature scenario. In this picture, the effective electronic temperature affects both the thermal distribution of the quasiparticles and the superconducting gap. The angle dependence of the photoexcited quasiparticle density and the quasiparticle decay rate are associated with the d-wave gap dynamics, while the fluence-dependent quasiparticle decay rate is attributed to the pairwise recombination of QPs into Cooper pairs. Different from the original explanations of the experimental results,\cite{sc1,sc3} the two-temperature scenario here doesn't refer to the details of the microscopic scattering processes, which have been washed out due to the thermalization of e-e scattering in the time scale $\tau_{ee}$. Our results suggest that the phenomenological two-temperature model could be a good  starting point to analyze the trARPES experiments before extracting other interesting microscopic dynamics processes. Furthermore, better time-resolution in experiment will be crucial for the development of more powerful trARPES technique.

\begin{acknowledgements}
We thank Jianqiao Meng for indicating us to the trARPES technique, and thank Weiqiang Chen, Zijian Yao, and Hongmin Jiang for helpful discussions. This work is supported by the Hong Kong grants of University Grant Council AoE/P-04/08 and GRC HKU707010, and by NSFC 11274269.
\end{acknowledgements}


\begin{thebibliography}{99}
\bibitem{TeEq} The original Rothwarf-Taylor equation for the total QP number $N$ in Ref.~\onlinecite{RT} reads $dN/dt=I_{0}+\beta N_{\omega}-RN^{2}$, where $I_{0}$ is the external source term to generate QPs, $N_{\omega}$ is the total number of phonon to break Cooper pairs with rate $\beta$, and $R$ is the recombination rate of QPs into Cooper pairs. In the two temperature model, we assume that the lattice temperature $T_{l}$ is unchanged and so is the phonon number $N_{\omega}$. Then the above equation for $N$ can be simplified as $dN/dt=-R(N^{2}-N_{l}^{2})$ if we neglected the source term $I_{0}$ and replaced the term $\beta N_{\omega}$ by $RN_{l}^{2}$ due to the equilibrium condition $dN_{l}/dt=0$. With the approximate relation $N=bT_{e}^{2}$ for d-wave superconductors, we get Eq.~(\ref{decay}) and the parameter $r=bR$ .
\end{thebibliography}
\end{document}